# Conversational Swarm Intelligence (CSI) Enables Rapid Group Insights

Louis Rosenberg
Unanimous AI
Pismo Beach, California
Louis@Unanimous.ai

Gregg Willcox
Unanimous AI
Seattle, Washington
Gregg@Unanimous.ai

Hans Schumann
Unanimous AI
San Francisco, California
Hans@Unanimous.ai

*Abstract*—When generating insights from human groups, conversational deliberation is a key method for exploring issues, surfacing ideas, debating options, and converging on solutions. Unfortunately, real-time conversations are difficult to scale, losing effectiveness in groups above 4 to 7 members. Conversational Swarm Intelligence (CSI) is a new technology that enables large human groups to hold real-time conversations using techniques modeled on the dynamics of biological swarms. Through a novel use of Large Language Models (LLMs), CSI enables real-time dialog among small groups while simultaneously fostering content propagation across a much larger group. This combines the benefits of small-scale deliberative reasoning and large-scale groupwise intelligence. In this study, we engage a group of 81 American voters from one political party in real-time deliberation using a CSI platform called Thinkscape. We then task the group with (a) forecasting which candidate from a set of options will achieve the most national support, and (b) indicating the specific reasons for this result. After only six minutes of deliberation, the group of 81 individuals converged on a selected candidate and surfaced over 400 reasons justifying various candidates, including 206 justifications that supported the selected candidate. We find that the selected candidate was significantly more supported by group members than the other options (p<0.001) and that this effect held even after six minutes of deliberation, demonstrating that CSI provides both the qualitative benefits of conversational focus groups and the quantitative benefits of largescale polling.

## I. INTRODUCTION

Real-time discussion is critical for human groups to generate and debate ideas, deliberate alternatives, prioritize options, and converge on mutually agreeable solutions. With the popularity of Enterprise Collaboration tools for text-based communication such as Slack and Discord, and for teleconferencing like Zoom and Microsoft Teams, it would appear that engaging large groups in thoughtful real-time dialog would be easier than in the past. Unfortunately, conversational deliberation is difficult to scale, rapidly losing effectiveness as groups grow beyond 5 to 7 members [12, 16 - 22]. As a consequence, those wanting to engage human groups sized for statistically significant samples for applications ranging from market research and political forecasting to participatory democracy, have little choice but to use polls, surveys, online forums, and other mechanisms that lose the interactive qualitative benefits of real-time deliberation.

In this paper, we describe and test a solution that is inspired by swarm-based methods in the field of Collective Intelligence.

In the field of Collective Intelligence (CI) researchers study how the input collected from large human groups can be processed to generate significantly more accurate output than individuals could produce on their own [1]. Common methods for harnessing the collective intelligence of large groups are based primarily on votes, polls, and surveys of various forms. A new CI method called Artificial Swarm Intelligence (ASI) has been developed in recent years and found to outperform common statistical methods in many scenarios, especially when populations harbor diverse or conflicting views [2-7]. Unlike traditional methods that aggregate statistically, ASI is based on the dynamics of biological swarms. It enables networked human groups to deliberate in real-time systems (i.e., *swarms*) and quickly converge on solutions they can best agree upon [2-7].

Biologists refer to the emergent decision-making abilities of natural systems as *Swarm Intelligence* because it enables many biological groups to function as "super-organisms" that can make more effective groupwise decisions than any individual member could make on their own [1, 8, 9]. ASI has been shown to enable networked human groups to achieve similar benefits, amplifying collective accuracy, insights, and cohesion through real-time dynamic interactions [10-11].

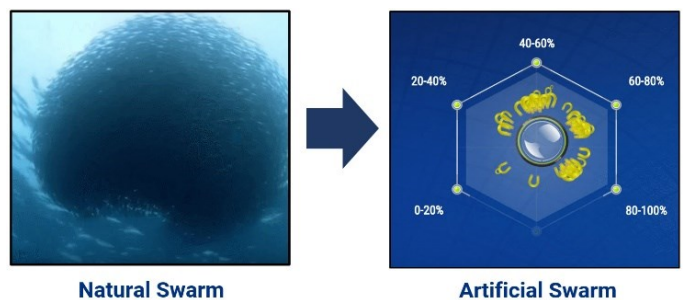

Fig. 1.  Natural Swarm vs ASI system of 100 participants

As shown in Figure 1 above, current methods for Artificial Swarm Intelligence involve use-cases where human groups deliberate among predefined sets of options and collectively rate, rank, or select among them. For example, ASI can be used for (a) forecasting the most likely outcome from a set of possible outcomes, (b) prioritizing sets of options into ranked lists that optimizes group satisfaction, or (c) rating the relative strengths of various options against specific metrics [4-6]. While these capabilities are useful in many real-world applications, there is a need for more flexible methods of deploying ASI, especially for enabling large human groups to engage in real-time conversational deliberation on open-ended questions in which a specific set of potential solutions is not known in advance.

In the following sections, we will describe an new approach for connecting networked human groups into real-time ASI systems, leveraging the intelligence amplification benefits of Swarm Intelligence via flexible conversational interactions.

## II. Conversational Swarm Intelligence (CSI)

To expand the applicability Artificial Swarm Intelligence to open-ended problems, a new architecture called Conversational Swarm Intelligence (CSI) has been developed and tested. The motivation for the CSI architecture is to enable large human groups to deliberate conversationally and converge on solutions that maximize group satisfaction, conviction, or accuracy. This poses a unique challenge for an online conversational system. For example, we could bring hundreds of people into a single chatroom but that would not yield meaningful dialog or insight. That's because conversational quality degrades with group size [12]. Sometimes referred to as the "many minds problem," when groups grow beyond a handful of people, the conversational dynamics fall apart, providing less "airtime" per person, disrupting turn-taking dynamics, providing less feedback per comment, and reducing engagement as participants feel less social pressure to participate. In fact, putting hundreds of people in a chatroom would not yield an authentic or meaningful "conversation" but instead would immediately devolve into a stream of singular comments with little interaction among them.

Fish schools, on the other hand, can hold "conversations" among hundreds or thousands of members with no central authority mediating the process. Each fish communicates with others using a unique organ called a "lateral line" that senses pressure changes caused by neighboring fish as they adjust speed and direction with varying levels of conviction. The number of neighbors that a given fish pays attention to varies from species to species, but it's always a small subset of the group. And because each fish reacts to an overlapping subset of other fish, information quickly propagates across the full population, enabling a single Swarm Intelligence to emerge that rapidly converges on unified decisions [13]. This is a powerful biological solution that was first emulated in networked human groups in 2021 using a technique called "hyperswarms" that was shown to enable real-time information propagation and solution convergence across a network of overlapping groups [14].

Researchers at Unanimous AI have recently applied this method to real-time human conversations via online chat. They developed an experimental platform called Thinkscape™ that is modeled on the communication dynamics of schooling fish. In this platform, a group of 400 online participants could be automatically divided into a large number of smaller subgroups, for example 80 groups of 5 people, with the members of each subgroup routed into their own chat room and tasked with discussing an issue in parallel with the other 79 subgroups. Merely subdividing into 80 groups does not yield an Artificial Swarm Intelligence of 400 people because information cannot propagate across the population. Researchers solved this issue through the novel use of AI agents powered by Large Language Models (LLMs) to emulate the functionality of the lateral line organ in fish.

To achieve this, an AI agent (referred to as an Observer Agent) is inserted into each of the 80 chat rooms and tasked with observing the deliberative dialog within that room, distilling the salient content, and expressing the content in a neighboring room through first-person dialog. In this way, each of the 80 groups is given an additional member that is an AI observer whose function is to conversationally convey (at intervals) the unique insights that emerge in one group into neighboring groups, thereby enabling information to propagate across the full population of 400 people.

This creates a single real-time system in which 400 or 4,000 or even 40,000 people could hold a real-time deliberation on a single issue, sharing thoughts and ideas, debating options, and converging on solutions that optimize overall support. In this way, CSI enables large groups to interact conversationally in real-time while ensuring that (a) individual participants can hold meaningful dialog in appropriately sized deliberative groups and (b) conversational content propagates globally leveraging the power of Artificial Swarm Intelligence. An example CSI structure for 100 members is shown below in Figure 2.

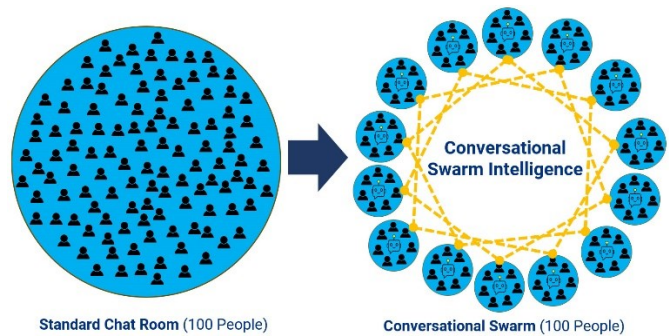

Fig. 2.  Standard Chat versus a Conversational Swarm for 100 members

Conversational Swarm Intelligence has two major benefits over prior ASI systems. First, it enables the use of open-ended questions in which answer options are not known in advance, empowering participants to suggest and debate an unlimited set of options that are not pre-defined. And second, it allows users to not only express which options they personally prefer but also argue for or against options with specific reasons. In this way, the Conversational Swarm Intelligence can not only facilitate convergence of large groups on unified solutions, it can quickly capture the full set of reasons why the group supports the particular solutions they converge on. In addition, the CSI structure is designed to mitigate social influence bias because (a) each individual is only influenced by a small number of others in real-time, and (b) ideas only propagate organically after they gain momentum locally [15].  In this way, particularly strong personalities (i.e., loudmouths) only have oversized influence in their small local group, greatly reducing their biasing effect.

For the reasons described above, CSI technology is designed to combine the quantitative benefits of conducting large-scale poll with the qualitative benefits of running intimate focus groups. Considering these unique advantages, we expect CSI to be useful for a wide range of applications from market research and organizational decision-making, to collaborative forecasting, political insights, and deliberative democracy. For example, CSI could be extremely useful for Deliberative Civic Engagement, enabling thoughtful conversations among samples of the public.

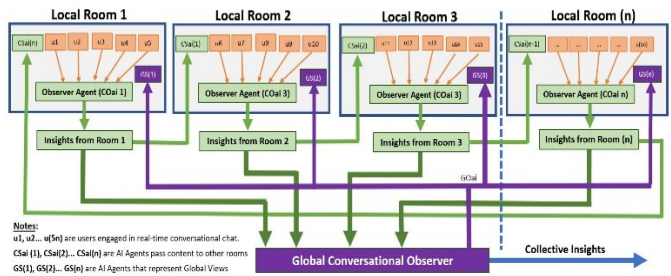

Fig. 3. Conversational Swarm Intelligence Architecture

As shown in figure 3 above, a novel architecture was developed to enable real-time conversational swarms among human users connected across standard computer networks. As shown, a full population (p) is broken up into (n) local rooms, each local room populated with approximately (p/n) users. Communication research suggests that ideal deliberative groups for real-time conversation range in size from 4 to 7 members, as studies show that conversations transition from authentic *dialog* to sequential *monologue* as groups approach 10 members [16]. Other research suggests maximum satisfaction for participants occurs around 5 members [17]. To optimize deliberative conversations in local groups, our current CSI heuristics aim to choose (n) such that (p/n) falls in the ideal range of 4 to 7 members.

Each local room is also assigned an Observer Agent (COai) driven by API calls to an LLM model, for example GPT-4. The AI agent was tasked with processing batches of conversational dialog within its assigned local room at intervals based on elapsed time and/or elapsed content. The processing step performs a set of functions on each block of observed dialog, which include – (i) identifying <u>newly proposed</u> suggestions related to the topic at hand, (ii) identifying reasons expressed <u>in support</u> of prior suggestions, (iii) identifying reasons expressed <u>in opposition</u> of prior suggestions, and (iv) estimating the conviction level when a member either suggests, supports, or opposes an alternative. In addition, the Observer Agent passes the summary and assessment to a databasing process that stores and aggregates suggestions and tracks groupwise conviction.

Each local room is also provided with a Surrogate Agent (CSai) tasked with expressing, at appropriate intervals, the local summaries, as captured from a neighboring room, into the room the agent is assigned to. In a prior study, the Thinkscape CSI system was tested at Carnegie Mellon University with real-time groups of 25 participants and compared to standard chat rooms. Even at this relatively small population size, these tests showed significant benefit, with participants in the CSI structure producing 30% more contributions (p<0.05) than those using a standard chat room and 7.2% less variance, indicating that users contributed more content and participated more evenly when using CSI [23].

In a second prior study of CSI, groups of 48 participants were tested in Thinkscape and tasked with debating the most significant risks that AI technology poses to human society within the next five years. The results revealed that when deliberating via CSI, the participants contributed 51% more content (p<0.001) as compared to participants using standard chat rooms. In addition, the deliberations using CSI showed 37% less difference in contribution quantity between the most vocal contributors vs the least vocal contributors, thereby indicating that CSI fosters more balanced deliberations. And finally, a large majority of the participants in this study preferred using CSI system over standard chat (p<0.05) and reported feeling more impactful when doing so (p<0.01) [24]

Based on these two prior studies of 25 person groups and 48 person groups, a new study of 80 person groups was conducted as described in Section III below. The objective of this new study is to assess the ability of CSI systems to generate and quantify insights, exploring how well large, distributed groups using CSI can deliberate on controversial topics, converge on solutions, and in the process generate reasons in support or opposition of various solutions. In this case, the controversial topic was the Republican Primary for President in 2023.

### III. INSIGHT STUDY

A cloud-based platform called Thinkscape was deployed using the CSI architecture described above. A pilot study was run to evaluate the effectiveness of conversational swarms in facilitating deliberative dialog and consensus judgements in a large, distributed group of 81 US Republican voters. The participants were asked to forecast which of the six leading Republican candidates who participated in a recent debate (Chris Christie, Ron DeSantis, Nikki Haley, Mike Pence, Vivek Ramaswamy, and Tim Scott) would the American public support most if they had to choose between only those six candidates. Participants were informed that they were working together to reach a conversational decision by real-time text chat and were shown a short video on how the CSI system worked.

Specifically, they were informed that the Thinkscape CSI system would divide them randomly into 15 parallel groups of approximately 5 or 6 individuals, with each group assigned to its own local chat room. They were also informed that each group would be assigned an Artificial Agent that would participate in their chat room as a conversational member, but would only perform the following functions: (a) observe their local conversation, (b) summarize their insights and occasionally pass those insights to a neighboring room in the networked structure, and (c) receive and conversationally express the insights from another room in the structure, thereby allowing this group to have access to the insights being discussed outside of their room. Although this may sound complex, a 90 second animation was used to make the structure and process very clear to participants.

Once instructed about the functionality of the CSI system, the 81 participants were automatically routed into their local rooms and tasked with deliberating together via free-flowing text chat with no imposed limitations as to when and how members contributed their conversational input during the six-minute period. The stated objective was for individuals to argue for or against various candidates, providing reasons why they held those positions, and collectively and reach a consensus for a single preferred answer within the short six-minute window. The consensus judgement was determined using a preference-labeling algorithm. This preference-labeling algorithm worked as follows: after every 5 messages in each room, or at most 15 seconds after a new message appeared in that room, API calls to an LLM (in this case, GPT 3.5) were automatically generated, causing the LLM to identify the answers that each person in that room preferred in response to the question being discussed. The strength of the individual preferences assessed were normalized on a scale of -3 (extreme negative) to +3 (extreme positive). This preference-labeling architecture was manually validated for accuracy of the preference labels and was deemed highly accurate at labeling user preference during real-time conversations in the ThinkScape platform. If a user did not mention an item or expressed no preference, their preference for that item was set to 0. The net preference for each item was calculated as the average preference for that item over all individuals. The item that showed the highest net preference at the end of the question was selected as the final answer.

## IV. Results

Within the allotted six minutes, the real-time deliberative conversation across 81 networked participants reached a clear and decisive consensus that Ron DeSantis (among the six candidates considered) was the most likely to be selected by the largest plurality of American voters. As captured in real-time by the Thinkscape system, over 200 statements in favor of DeSantis were expressed, noting for example that he has high name recognition and is currently a popular governor of a large state. Arguments in favor of DeSantis also cited his appeal to independents and young voters. Participants who argued against DeSantis, however, expressed that he lacks appeal outside of his own party, is a highly polarizing leader, and is viewed by many as an anti-science figure. DeSantis' track record in Florida schools was a divisive issue, with some participants citing his record as a primary reason for favoring him while others expressed it would hurt his chances for broad support. For reference, the verbatim LLM generative summary output by Thinkscape is provided below for candidate Ron DeSantis:

Arguments in Favor of Ron DeSantis:

*"62 participants argued in favor of Ron DeSantis because they believe that DeSantis has high name recognition and is the popular governor of a big state, which gives him an advantage. They also mention his representation of younger generations and the belief that he would receive a lot of right-leaning votes by default. Additionally, they highlight his upvotes by Elon Musk, his tough image, and his experience in handling COVID-19 lockdowns. Other reasons include his support for education and opposition to woke principles, his popularity and media presence, and the belief that he has the best chance of winning the primary and competing against Biden. They also mention his experience as governor of Florida, his national appeal, and his large donor base. Overall, these group members believe that DeSantis is the most electable candidate with broad appeal to independents and a strong track record as a successful governor."*

Arguments Against Ron DeSantis:

*"24 participants argued against Ron DeSantis because they believe he is scandalous, lacks the ability to engage in retail politics, is too close to Trump, is polarizing, and has a negative track record in Florida, particularly in regard to the education system. They also criticize DeSantis for being anti-science, hiring a questionable state Surgeon General, and discouraging vaccination. The group members argue that DeSantis is too radical and polarizing for non-Republicans, and that he lacks appeal outside of his own party. They believe that he needs to move towards the center and that there are too many extremists currently having their way. Additionally, they mention that there are many people in Florida who dislike DeSantis, and they question his ability to win outside of the Republican base. Some members express personal dislike for DeSantis, questioning his character and suggesting he may be an operative from the other side. Overall, they believe that DeSantis is damaged goods, losing steam, and not the right candidate to win the support of non-Republicans and moderates."*

The specific points aside, what is possibly most interesting is the thoughtful breadth of insights generated in only six minutes by 81 networked individuals holding 15 parallel conversations with information propagation driven by conversational AI agents. Furthermore, similarly detailed arguments were expressed, captured, and summarized for each of the other five candidates, all within the same six-minute window. Thus, it seems that the CSI structure enabled highly efficient deliberation and insight across 80+ participants, avoiding the sluggish turn-taking dynamics that occur in groups larger than 4 to 7 people.

Digging deeper, we can quantify the volume of arguments that were conversationally generated. Table I details the number of reasons (for and against each candidate) that were expressed during the short CSI conversation. A total of 410 reasons were verbalized in the six-minute session (266 in favor of candidates, 144 against), an average of 5 per participant. Ron DeSantis was by far the most talked-about candidate, with a total of 206 reasons expressed in favor and 54 against. Nikki Haley had the next-most reasons expressed in favor (28) and was the only other candidate to have significantly more reasons expressed in favor than expressed against.

TABLE I. Number of Reasons Surfaced in Favor and Against Each Candidate

| Candidate | Number of Reasons in Favor | Number of Reasons Against |
|---|---|---|
| Ron DeSantis | 206 | 54 |
| Nikki Haley | 28 | 19 |
| Mike Pence | 16 | 30 |
| Vivek Ramaswamy | 11 | 35 |
| Chris Christie | 3 | 6 |
| Tim Scott | 2 | 0 |
| **Total** | **266** | **144** |

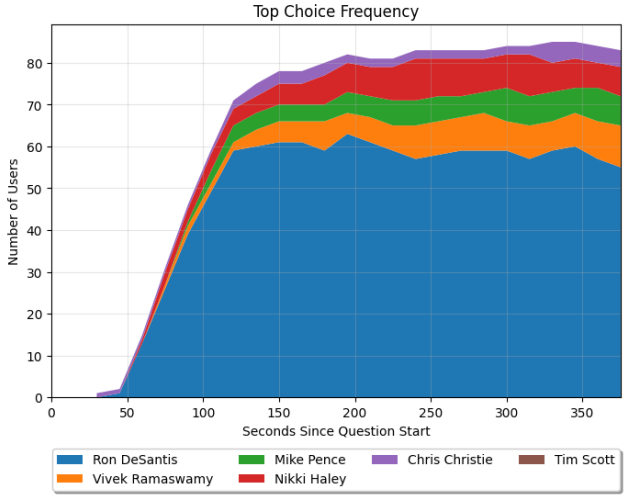

Fig. 4. Frequency of User Top Choices: About 70% of users supported Ron DeSantis throughout the question.

Using the preference-labeling architecture described above, each user's top-choice was identified at multiple intervals through the deliberation. These top choices were tallied and are shown in Figure 4 above as a stacked line plot. Of the total 81 unique users, approximately 60 (74%) users supported DeSantis from the outset of the question, declining slightly to 55 (68%) by the end of the question. So, only a few users were convinced to switch their top choice from DeSantis to other answers. Vivek Ramaswamy, Nikki Haley, and Mike Pence all grew to similar shares over the deliberation, ending at 10, 7, and 6 users respectively. The remaining users supported Chris Christie.

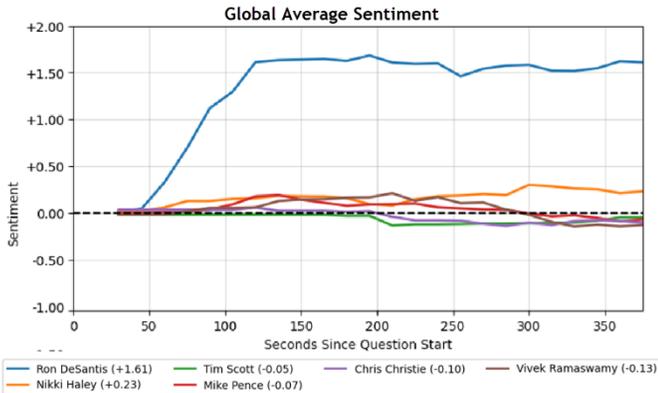

Fig. 5. Frequency of User Top Choices: About 70% of users supported Ron DeSantis throughout the question.

The average group sentiment towards each candidate over time is shown in Figure 5 above. Ron DeSantis grew a commanding lead in the first two minutes of conversation and retained this lead throughout the discussion, ultimately resulting in DeSantis being the selected candidate. To evaluate whether the group reached the final answer of DeSantis by random chance, a paired t-test was conducted that compared the average sentiment of DeSantis to each other answer. The tests were run after each of the three time periods in the session, as shown in Table II: Initialization (0 to 150 sec), Deliberation (150 to 300 sec), and Convergence (300 to 400 sec). DeSantis was always the candidate with the highest sentiment by a margin of at least 1.3 points, and at each time period had a significantly higher average sentiment across all candidates and all time periods ($p<0.01$).

TABLE II. AVERAGE SENTIMENT OF EACH CANDIDATE BY TIME PERIOD.

| Candidate | Time Period | | |
|---|---|---|---|
| | Initialization | Deliberation | Convergence |
| Ron DeSantis | 1.81 | 1.62 | 1.67 |
| Nikki Haley | 0.20** | 0.31** | 0.24** |
| Mike Pence | 0.16** | -0.01** | -0.07** |
| Vivek Ramaswamy | 0.16** | -0.02** | -0.13** |
| Chris Christie | 0.03** | -0.11** | -0.11** |
| Tim Scott | -0.02** | -0.11** | -0.05** |

** indicates significant result at p=0.01 level.

This result indicates that DeSantis was seen as the candidate most likely to garner broad support, not only at the outset of the conversation, but also after a long period of deliberation during which alternate arguments were expressed. In other words, DeSantis was both the top-of-mind choice and held up under criticism from users who favored other candidates.

## V. CONCLUSIONS

We test Conversational Swarm Intelligence, a novel LLM-based technology that enables large, distributed groups to hold coherent deliberations in real-time via online chat and reach thoughtful and unified consensus. We performed a study with a large 81-member group of Republicans, testing their ability to rapidly discuss a controversial topic and converge on a solution while generating a large set of open-ended insights in support and opposition of the chosen answer.

In this study, the 81-member group quickly converged on the collaborative assessment that Ron DeSantis was the Republican candidate that is most likely to be chosen by a representative sample of Americans. A total of 410 reasons (for and against various candidates) were expressed and identified over the six-minute session, the majority of which supported Ron DeSantis as the best answer to the question. A majority (68%) of users supported DeSantis at the end of the question, and the group's preference for DeSantis was shown to be statistically higher ($p<0.01$) than their preference for any answer choice throughout each stage of the discussion: Initialization, Deliberation, and Convergence.

This work serves as a proof-of-concept that large groups can engage in deliberations with CSI and that both quantitative and qualitative insights can be extracted from real-time freeform CSI discussions, paving the way for the scalable collection of conversational insights from human populations. Future work will explore whether CSI improves the accuracy of collective decision-making, the effectiveness of groupwise brainstorming, and representativeness of collective assessments. Future work will also strive to test CSI in real-time deliberative democracy and civic engagement contexts. And finally, future work will test substantially larger groups, enabling real-time deliberations among many hundreds or thousands of simultaneous users. The goal of developing largescale CSI is to explore methods for achieving new levels of groupwise intellect, potentially even unlocking a pathway for reaching Collective Superintelligence.

ACKNOWLEDGMENT

The authors thank Chris Hornbostel and Patty Sullivan for their efforts recruiting participants and moderating sessions.

REFERENCES

[1] Malone T., Superminds: The surprising power of people and computers thinking together. London: Oneworld Publications, 2018. 374 pp, paperback. ISBN: 978 1 78607 470 6

[2] Rosenberg, L., Human Swarms, a real-time method for collective intelligence. In: Proceedings of the European Conference on Artificial Life 2015, ECAL 2015, pp. 658–659, York, UK. MIT Press (2015). ISBN 978-0-262-33027-5. https://doi.org/10.1162/978-0-262-33027-5-ch117

[3] Rosenberg, L.: Artificial Swarm Intelligence, a human-in-the-loop approach to A.I. In: Proceedings of the Thirtieth AAAI Conference on Artificial Intelligence, AAAI 2016, Phoenix, Arizona, pp. 4381–4382. AAAI Press (2016) https://doi.org/10.1609/aaai.v30i1.9833

[4] Metcalf, L., Askay, D. A., & Rosenberg, L. B. (2019). Keeping Humans in the Loop: Pooling Knowledge through Artificial Swarm Intelligence to Improve Business Decision Making. California Management Review, 61(4), 84–109.

[5] Patel, B.N., Rosenberg, L., Willcox, G. et al. Human–machine partnership with artificial intelligence for chest radiograph diagnosis. Nature npj Digit. Med. 2, 111 (2019).

[6] Rosenberg, L., Willcox, G. (2020). Artificial Swarm Intelligence. In: Bi, Y., Bhatia, R., Kapoor, S. (eds) Intelligent Systems and Applications. IntelliSys 2019. Advances in Intelligent Systems & Computing, vol 1037. Springer, Cham.

[7] Schumann, H., Willcox, G., Rosenberg, L. and Pescetelli, N., "Human Swarming Amplifies Accuracy and ROI when Forecasting Financial Markets," 2019 IEEE International Conference on Humanized Computing and Communication (HCC 2019), Laguna Hills, CA, USA, 2019, pp. 77-82 10.1109/HCC46620.2019.00019

[8] Seeley, Thomas D., et al. "Stop signals provide cross inhibition in collective decision-making by honeybee swarms." Science 335.6064(2012): 108-111.

[9] Seeley, Thomas D. Honeybee Democracy. Princeton Univ. Press, 2010

[10] L. Rosenberg, M. Lungren, S. Halabi, G. Willcox, D. Baltaxe and M. Lyons, "Artificial Swarm Intelligence employed to Amplify Diagnostic Accuracy in Radiology", 2018 IEEE 9th Annual Information Technology Electronics and Mobile Communication Conference (IEMCON), pp. 1186-1191, 2018 .

[11] Askay, D., Metcalf, L., Rosenberg, L., Willcox, D.: Enhancing group social perceptiveness through a swarm-based decision-making platform. In: Proceedings of 52nd Hawaii International Conference on System Sciences, HICSS-52. IEEE (2019)

[12] Cooney, G., et. al., "The Many Minds Problem: Disclosure in Dyadic vs. Group Conversation." Special Issue on Privacy and Disclosure, Online and in Social Interactions edited by L. John, D. Tamir, M. Slepian. Current Opinion in Psychology 31 (February 2020): 22–27.

[13] Parrish, J. K., Viscido, S. and Grünbaum, D., "Self-Organized Fish Schools: An Examination of Emergent Properties." Biological Bulletin 202, no. 3 (2002): 296–305.

[14] Willcox, G., Rosenberg, L., Domnauer, C. and Schumann, H., "Hyperswarms: A New Architecture for Amplifying Collective Intelligence," 2021 IEEE 12th Annual Information Technology, Electronics and Mobile Communication Conference (IEMCON), Vancouver, BC, Canada, 2021, pp. 0858-0864, doi: 10.1109/IEMCON53756.2021.9623239.

[15] Muchnik, L. et al., Social Influence Bias: A Randomized Experiment. Science 341, 647-651 (2013). DOI:10.1126/science.1240466

[16] Fay, N., Garrod, S., & Carletta, J. (2000). Group Discussion as Interactive Dialogue or as Serial Monologue: The Influence of Group Size. Psychological Science, 11(6), 481–486. https://doi.org/10.1111/1467-9280.00292

[17] Hackman, J. Richard, and Neil Vidmar. "Effects of Size and Task Type on Group Performance and Member Reactions." Sociometry 33, no. 1 (1970): 37–54. https://doi.org/10.2307/2786271.

[18] Nabatchi, Tina. (2012). An Introduction to Deliberative Civic Engagement. 10.1093/acprof:oso/9780199899265.003.0001.

[19] McCoy, M., Scully, P. (2002). Deliberative Dialogue to Expand Civic Engagement: What Kind of Talk Does Democracy Need? National Civic Review, vol. 91, no. 2, Summer 2002

[20] Nabatchi, T. (2014) "Deliberative Civic Engagement in Public Administration and Policy," Journal of Public Deliberation: Vol. 10: Issue 1, Article 21

[21] Gastil, J., & Levine, P. (2005). The deliberative democracy handbook: Strategies for effective civic engagement in the twenty-first century. Jossey-Bass.

[22] Nabatchi, T., & Leighninger, M. (2015). Public participation for 21st century democracy. John Wiley & Sons.

[23] Rosenberg, L., et. al.,"Conversational Swarm Intelligence, a Pilot Study." arXiv.org, August 31, 2023. https://arxiv.org/abs/2309.03220.

[24] Rosenberg, L., Willcox, G., Schumann, H. and Mani, G., "Conversational Swarm Intelligence (CSI) Enhances Groupwise Deliberation." 7th International Joint Conference on Advances in Computational Intelligence (IJCACI 2023). Oct 14, 2023. New Deli.